\documentclass[%
 reprint,
nofootinbib,
 amsmath,amssymb,
 prx
]{revtex4-2}

\usepackage{graphicx}
\usepackage{dcolumn}
\usepackage{bm}
\usepackage[dvipsnames]{xcolor}
\usepackage{natbib}

\begin{document}

\preprint{APS/123-QED}

\title{Viability of quantum communication across interstellar distances}

\author{Arjun Berera}
 \author{Jaime Calder\'on-Figueroa}%
\affiliation{%
 School of Physics and Astronomy, University of Edinburgh, Edinburgh EH9 3FD, United Kingdom
}%


\begin{abstract}
The possibility of achieving quantum communication using photons across interstellar distances is examined. For this, different factors are considered that could induce decoherence of photons, including the gravitational field of astrophysical bodies, the particle content in the interstellar medium, and the more local environment of the Solar System. The X-ray region of the spectrum is identified as the prime candidate to establish a quantum communication channel, although the optical and microwave bands could also enable communication across large distances. Finally, we discuss what could be expected from a quantum signal emitted by an extraterrestrial civilization, as well as the challenges for the receiver end of the channel to identify and interpret such signals.  
\end{abstract}

                              
\maketitle
{\it In Press Physical Review D 2022}
\section{Introduction}

The most significant technological revolution of the last decades has arguably been the creation and assimilation of the internet globally. This innovation in the way we communicate has been possible due to the deployment a myriad of classical channels and technologies that exploit the properties of electromagnetic radiation. Indeed, our telecommunications are based on the transmission and reception of signals that travel as electromagnetic waves in free space, or as photonic pulses in optical fiber. In these processes, we encode our messages on physical features such as amplitudes, frequencies, voltages, and so on \cite{mishra2016}. However, the use of the quantum mechanical features of light (or any other particle for that matter) has opened the door for more efficient communication protocols and information processing \cite{Nch10,zeng2010quantum,Krenn2016}. 
Quantum phenomena are also quite useful for encryption purposes, spinning off an area known as quantum cryptography. One of its main applications, quantum key distribution (QKD), allows observers to detect the presence of any eavesdropper \cite{bennett2020quantum,grosshans2002continuous,inoue2003differential,sasaki2014practical}.

The perks of quantum communication, and the development of technology that can realize it at ever-increasing distances, have inspired the study of this process not only at global scales, but also for near Earth space communications. There have been many theoretical and experimental studies looking at the establishment of a quantum communication channel between a laboratory on Earth, and a satellite at a low Earth orbit (LEO) \cite{bourgoin2013comprehensive,yin2012quantum,yin2013experimental,carrasco2016leo,liao2017satellite,yin2017satellite,noh2021quantum,Kohlrus:2018cnb}. Under such circumstances, it has been clear there is a necessity to understand how relativistic effects can alter the communication protocols developed for terrestrial communications. Moreover, one of us pointed out that it is possible to go even further in terms of distances since the quantum state of a photon at certain frequencies can avoid decoherence to galactic scales \cite{Berera:2020rpl}. As such, the novel suggestion was made that, in principle, it should be possible to detect a quantum signal coming from an astrophysical body or even an intelligent signal from an extraterrestrial (ET) civilization. The rapid technological growth in the field of quantum communication is bringing into the realm of possibility the search for such signals. 

In this paper, we explore some of the potential factors that could prevent us from detecting quantum signals (or establishing quantum communication channels) at interstellar scales. In previous works, we have established that factors like the expansion of the Universe or the interaction with CMB radiation and other particles at galactic scales do not destroy the quantum coherence of a photon state \cite{Berera:2020rpl, Berera:2021xqa}. Here, we will pay attention to the potential role of gravity as a decoherence factor for photons. For this, we will review the concept of Wigner rotation and its relation to decoherence. In addition, we will also look at more ``environmental'' factors, like the potential disruption of the state due to interactions with the photon population coming from the Sun.

It is relevant to mention that, considering the kind of questions we are asking, technically, we can only determine how unlikely is decoherence at any astronomical scale. Even if the quantum coherent nature of the photon state is not destroyed, it is quite possible for the fidelity of the signal to be impaired. Indeed, decoherence implies the classicalization of a state through the interaction of a physical system with an environment, effectively producing the collapse of the wavefunction. On the other hand, fidelity is a measure of how close is the received state in comparison to the one that is sent. Thus, a successful establishment of a quantum communication channel requires not only the lack of decoherence of the photon state but also a `healthy' fidelity. We shall quantify what this means in Section \ref{s3}, where we also illustrate in more detail the difference between decoherence and fidelity by looking at particular examples. With this in mind, any practical search for extraterrestrial signals involving quantum communication would require a fair amount of guesswork. We will comment on this topic while also discussing the kind of signals that can be expected from an ET trying to establish contact through quantum teleportation and quantum communication in general. Finally, we argue why quantum communication could be the preferred mean of contact to the detriment of classical channels. 

\subsection{On the quantum state of photons}
Before going into the details of quantum communication through photons and what factors can disrupt it in space, let us quickly review a few basic but crucial aspects of the quantum theory of electromagnetism, which ultimately leads to the concept of photon. We will also comment on why one can assign a quantum state a la Schr\"odinger despite working on a manifestly field theory framework. 

The concept of photon is a direct consequence of the quantum nature of light, an idea first explored by Einstein in order to explain the photoelectric effect \cite{Einstein1905}. However, it was Dirac who described radiation as the simultaneous manifestation of the wave and particle duality of light. In this way, phenomena like diffraction and interference, which had been successfully described by a wave theory of light, could be understood while also looking at the excitations of an atom located along a wave front as the absorption of a quanta of energy of the field \cite{Dirac:1927dy}. Thus, Einstein's photons could be associated to quantized simple harmonic oscillators, which in turn are associated to the modes of the electromagnetic field. 
Let us quickly show the quantization of the (free) electromagnetic field, following \cite{mandel95, *cohen97}. First, we expand the magnetic vector potential ${\bf A}$ in its Fourier modes,
\begin{equation}
    \bf{A}({\bf r},t) = \frac{1}{\sqrt{\epsilon_0 V}} \sum_{\bf k} {\bf {\cal A}}_{\bf k} (t) e^{i {\bf k} \cdot {\bf r}}\;.
\end{equation}
Choosing the transverse gauge, we have
\begin{equation}
    \frac{i}{\sqrt{\epsilon_0 V}} \sum_{\bf k} {\bf k} \cdot {\bf {\cal A}}_{\bf k} (t) e^{i {\bf k} \cdot {\bf r}} = 0 \implies {\bf k} \cdot {\bf {\cal A}}_{\bf k} (t) = 0\;,
\end{equation}
which ultimately leads to the reality condition ${\bf {\cal A}}_{-\bf k} (t) = {\bf {\cal A}}_{\bf k}^* (t)$. Then, the Fourier modes of the field satisfy the wave equation
\begin{equation}
    \left(\frac{\partial^2}{\partial t^2} + \omega^2 \right) {\bf {\cal A}}_{\bf k} (t) = 0\;,
\end{equation}
where $\omega = kc$, and which has solutions of the form
\begin{equation}
    {\bf {\cal A}}_{{\bf k}} (t) = \sum_s \left[ {\cal E}_{{\bf k}s} a_{\bf k} e^{-i \omega t} + {\cal E}_{-{\bf k}s}^* a_{- \bf k}^* e^{i \omega t} \right]\;,
\end{equation}
where ${\cal E}_{{\bf k}s}$ denotes the two possible orthogonal polarization vectors. It follows that the electric and magnetic fields can be respectively written as
\begin{equation}
    {\bf E} ({\bf r},t) = \frac{i}{\sqrt{\epsilon_0 V}} \sum_{{\bf k},s} \omega [u_{{\bf k}s}(t) {\cal E}_{{\bf k}s} e^{i {\bf k}\cdot{\bf r}} - {\rm c.c.}]\;,
\end{equation}
\begin{equation}
     {\bf B} ({\bf r},t) = \frac{i}{\sqrt{\epsilon_0 V}} \sum_{{\bf k},s} \omega [u_{{\bf k}s}(t) ({\bf k} \times {\cal E}_{{\bf k}s}) e^{i {\bf k}\cdot{\bf r}} - {\rm c.c.}]\;,
\end{equation}
where for future convenience we have introduced $u_{{\bf k}s}(t) = a_{{\bf k}s} e^{-i \omega t}$. With these expressions at hand, we can write the energy of the field as
\begin{align}
    H & = \frac{1}{2} \int \left[\epsilon_0 {\bf E}^2 ({\bf r},t) + \frac{1}{\mu_0} {\bf B}^2 ({\bf r},t) \right] dV \nonumber \\
    & = 2 \sum_{{\bf k},s} \omega^2 | u_{{\bf k}s} (t) |^2\;.
\end{align}
As is well known, the quantization of the field requires to promote the fields to operators, and to impose commutation relations on its modes (or alternatively on canonical variables which in turn imply the commutation relations of the Fourier components). These relations are
\begin{equation}
    \left[ \hat{a}_{{\bf k}s}(t), \hat{a}_{{\bf k'}s'}^{\dagger}(t)\right] = \delta^3_{{\bf kk'}} \delta_{ss'}\;,
\end{equation}
with any other combination yielding zero.\footnote{We require to rescale the Fourier components as $\hat{a}_{{\bf k}s} \rightarrow \sqrt{\frac{\hbar}{2 \omega \epsilon_0 V}} \hat{a}_{{\bf k}s}$, and similarly for $\hat{a}_{{\bf k}s}^{\dagger}$.} Then, the Hamiltonian operator is given by  
\begin{equation}
    \hat{H} = \sum_{{\bf k},s} \hbar \omega \left[\hat{a}^{\dagger}_{{\bf k}s} (t) \hat{a}_{{\bf k}s} (t) + \frac{1}{2} \right]\;,
\end{equation}
so that it has the same eigenstates of the number operator $\hat{n}_{{\bf k}s} = \hat{a}^{\dagger}_{{\bf k}s} \hat{a}_{{\bf k}s}$. Then, one can show how $\hat{a}^{\dagger}_{{\bf k}s}$ creates one quanta of a state with energy $\omega$ and helicity $s$, whilst $\hat{a}_{{\bf k}s}$ annihilates it. Accounting for every possible mode, a general state vector can be expressed as
\begin{equation}
    |\psi \rangle = \sum_{\{n_{\bf k}\}} c_{\{ n_{\bf k} \}} | \{ n _{\bf k} \}\rangle \;,
\end{equation}
where $\{ n _{\bf k} \}$ denotes the set of occupation numbers of every mode. Notice that in principle this should not be interpreted as a wavefunction of the photon(s). Heuristically, this can be understood due to the fact that a wavefunction can be projected into position space and we can ask questions such as the probability to find a (matter) particle in a certain position. It is not sensible to ask such questions for light, and instead questions such as the probability of the photon, characterized by the state $|\psi\rangle$, to excite an atom at a certain positions can be addressed. However, it can be useful to work with localized excitations of the electromagnetic field, especially in a laboratory setting or, for example, for quantum communication processes. Even though one must be careful with the interpretation of this `wavefunction', we can introduce a state like the following
\begin{equation}
    | \psi \rangle = C \sum_{\bf k} e^{-({\bf k}-{\bf k_0})^2/4\sigma^2} e^{-i {\bf k} \cdot {\bf r_0}} |1_{{\bf k}s},\{0\} \rangle\;,
\end{equation}
which is a linear superposition of one-photon states. Notice how this state does not have a definite momentum, and can be instead interpreted as a Gaussian wave packet spread about ${\bf k_0}$. With this caveat in mind, we can work with single-photon states a la Schr\"odinger, as we will do in the next section. 

\section{Quantum teleportation}
\label{qtssect}
Quantum communication, as a field, is closely related to quantum information processing and quantum teleportation, with quantum entanglement as a common thread between them \cite{Bennett:1992tv,bennett1992communication,ekert1992quantum}. 
Indeed, the entanglement of a quantum state can be utilized in various aspects of treating information, and it allows for the information of such state to be transferred from one location to another. Let us illustrate this by considering an observer, Alice ($A$), who is in possession of a quantum state $| \chi \rangle $ characterizing a system like a spin-1/2 particle, a photon or some such identical particle. Alice wants to send the complete information of the quantum state $| \chi \rangle$ (which need not be known to her) to Bob ($B$), who also has one of these identical particles. Through quantum teleportation, the particle at Bob's location can be placed in exactly the same state $|\chi\rangle$ as the particle Alice was in possession of. In doing so, it is as if the material system in one location has been moved to another location. However, quantum teleportation does not physically move particles, but rather it is a process of communication requiring both a shared entanglement and a classical channel. As one of the components involves classical communication, information cannot be transferred any faster than the speed of light \cite{Bennett:1992tv}.

A simple example of quantum teleportation
is as follows. Consider two photons 
prepared in the Bell state, 
\begin{equation}
|\Psi_{AB}^- \rangle = \frac{1}{\sqrt{2}} \left[|+_A -_B \rangle - |-_A +_B \rangle \right] \;,
\label{psiminus}
\end{equation}
where $| + \rangle$ and $| - \rangle$ denote the two polarization states of the photon, and the subscripts label who is in possession of the respective state. It is crucial that Alice and Bob agree beforehand on the above entangled state, regardless of how it was prepared or handed over to them. Setting this up is known as establishing a shared entanglement. Next, Alice has an additional photon in the (normalized) state
\begin{equation}
    | \chi_{A'} \rangle = c |+_{A'} \rangle + d | -_{A'} \rangle \;,
\end{equation}
whose information she wishes to send to Bob. Then, the three-particle state is
\begin{eqnarray}
|\Phi_{A'AB} \rangle & = & c |+_{A'} \rangle |\Psi^-_{AB} \rangle + d |-_{A'} \rangle |\Psi^-_{AB} \rangle
\\
& = & \frac{c}{{\sqrt{2}}} \left[ |+_{A'} \rangle |+_A \rangle |-_B \rangle
- |+_{A'} \rangle |-_A \rangle |+_B \rangle \right]
\nonumber \\
& + &
\frac{d}{{\sqrt{2}}} \left[ |-_{A'} \rangle |+_A \rangle |-_B \rangle
- |-_{A'} \rangle |-_A \rangle |+_B \rangle \right] 
\nonumber \;.
\label{phiaab}
\end{eqnarray}
This expression can be rewritten in terms of the Bell states basis\footnote{The Bell states are $| \Psi^{\pm} \rangle = \frac{1}{\sqrt{2}} ( |+ - \rangle \pm |- + \rangle)$ and $| \Phi ^{\pm} \rangle = \frac{1}{\sqrt{2}} ( |+ + \rangle \pm |- - \rangle) $. They are referred to as one \emph{ebit} of entanglement.} for Alice as follows
\begin{align}
|\Phi_{A'AB} \rangle = & \frac{1}{2} \big[
|\Psi^-_{A'A} \rangle (-c | +_B \rangle - d |-_B \rangle) 
 +|\Psi^+_{A'A} \rangle (-c | +_B \rangle \nonumber \\
 & + d |-_B \rangle) + |\Phi^-_{A'A} \rangle (c | -_B \rangle + d |+_B \rangle) \nonumber \\
 & +|\Phi^+_{A'A} \rangle (c | -_B \rangle - d |+_B \rangle)
\big] \;.
\label{bellform1}
\end{align}
We see that in this basis the states at $B$ are related to $| \chi \rangle$ by unitary transformations. Then, if Alice makes an observation of one of the above four Bell states, she can communicate the outcome to Bob via a classical channel, requiring two classical bits of information. That is the crucial fact that ensures that information cannot travel faster than light. Moreover, once Alice performs an observation (effectively teleporting the state $| \chi \rangle $ to Bob), the state  
$| \chi_{A'} \rangle$
at her side is destroyed, in agreement with the no-cloning theorem.

The above has been an idealized description of quantum teleportation. There are several technical problems that need to be addressed to actually realize such a process.  First, how is the entangled Bell state created. For this there are methods such as spontaneous parametric down conversion \cite{zukowski1993event,shapiro2002architectures,clausen2014source}. Then, having created
a Bell state, both particles of this state need to be given to the two respective observers $A$ and $B$.  If there are any environmental factors that interfere with either of the two particles in the transmitting or receiving apparatus or on their journey to the two observers, the quantum coherence between the two particles will of course be degraded \cite{shapiro2002architectures,bourgoin2013comprehensive}. This will be an important point to consider for demands placed on the ultra-long distances required for interstellar teleportation and will be discussed in the following sections.  What is known from experiments is quantum teleportation has been successful over long distances with photons propagating both through fiber optic cables \cite{marcikic2003long,ursin2004quantum} up to distances of order 10 km, through the atmosphere close to sea level, up to around a hundred kilometers \cite{boschi1998experimental,fedrizzi2009high,yin2012quantum}, and up to over a thousand kilometers via satellite to ground teleportation \cite{yin2017satellite}. Both through fiber optics and low altitude atmospheric transmission, loss of signal to the medium has limited the distance to which teleportation can be successful.  For the satellite based experiment, where most of the photon journey remains at altitudes above $\sim 10\ {\rm km}$, attenuation loss
is substantially reduced, thus allowing for much longer distance teleportation. Another detail
is that a Bell state measurement has to be performed by the observer $A$ who wants to send the quantum state to $B$. Various techniques have been developed to make such measurements
\cite{bouwmeester1997experimental,pirandola2015advances}.

\section{Gravitational effects on the photon quantum state}\label{s3}

A priori, one of the ways that the quantum state of a photon could be disrupted is through the interaction with a gravitational field. Implementing a quantum communication channel at astronomical scales implies that photons may need to travel across the gravitational field generated by many astrophysical sources. Naturally, that would be the case if either the transmitter or the receiver are on a planet or in an orbit near one. Another common source to account for are nearby stars, such as for solar--system type environments.

The study of gravitational effects on quantum communication is not new. Primarily, the problem of establishing quantum communication with a satellite in a low Earth orbit has been tackled. In one way or another, the main goal is to compute the \emph{fidelity} of the channel and to determine how gravity affects it. However, here we are contemplating the idea of being just on the receiver end of the communication channel, without any control over the protocol or any other factor. As such, as a first step, one can only ask about the feasibility of photons avoiding decoherence. We will argue in favor of this point while stressing that lack of decoherence is not equivalent to maintaining full fidelity. Thus, even if such signals exist, a big part of the detection and processing would be the guesswork involved in it. In this way, the lack of decoherence is considered a necessary starting point.

In order to make our case for the lack of decoherence for massless
$U(1)$ bosons (photons), we shall briefly review the concept of Wigner rotation, 
first looking more generally at how the states corresponding to 
both massive and massless particles transform under unitary operators and the crucial difference between them. Next, we will look at a formalism to compute the fidelity specifically for photons under gravitational influence. We will see that even though the fidelity can change, one can achieve relatively minor corrections to the signal for certain cases, providing further evidence that an interstellar contact through quantum communication could be achieved.

\subsection{Wigner rotation}

The concept of Wigner rotation emerges in the context of the representations of the Poincar\'e group (translations + rotations + boosts in flat spacetime). The Poincar\'e group is noncompact, so all its unitary representations are infinite-dimensional. Nevertheless, Wigner managed to classify the states by looking instead to the representations of a suitable subgroup of the Lorentz group, known as  the \emph{little group} \cite{Wigner:1939cj}. 

To see how this comes about, let us start with a single-particle state $|p, \sigma \rangle$, where $p$ denotes its momentum and $\sigma$ any other possible labels, which are ultimately determined by the Casimir invariants of the group. Then, we have that 
\begin{equation}
     P^{\mu} | p, \sigma \rangle = p^{\mu} | p, \sigma \rangle\;,
\end{equation}
where $P^{\mu}$ is the momentum operator. Now, let $\Lambda$ be a Poincar\'e transformation and $U(\Lambda)$ the induced unitary linear transformation in a Hilbert space, such that $|\Psi \rangle \rightarrow U(\Lambda) |\Psi\rangle$. One can then show that $U(\Lambda)|p, \sigma \rangle$ is an eigenstate of $P^{\mu}$ with eigenvalue $\Lambda p$. Hence, it follows that
\begin{equation}\label{uL}
    U(\Lambda) |p, \sigma \rangle = \sum_{\sigma'} C_{\sigma \sigma'} (\Lambda, p) |\Lambda p, \sigma' \rangle
    \;,
\end{equation}
where, by an appropriate choice of basis, $C_{\sigma \sigma'}$ can be made block-diagonal, and one could look at them to arrive to the irreducible representations \cite{Weinberg:1995mt,Sundermeyer:2014kha}.

Notice that the states in the space spanned by the eigenvectors $|p, \sigma \rangle$ -- with $p^2$ the same for all of them -- are isomorphic to each other \cite{Costa:2012zz}. Then, we can take a fixed fiducial four-vector $k$, for which there exists a Lorentz transformation such that $ p^{\mu} = L^{\mu}{}_{\nu} k ^{\nu}$, and 
\begin{eqnarray}\label{ul}
    |p, \sigma \rangle = N(p) U(L(p)) |k, \sigma \rangle\;,
\end{eqnarray}
where $N(p)$ is a normalization factor. From the eqs. (\ref{uL}) and (\ref{ul}), we have
\begin{equation}
    U(\Lambda) | p, \sigma \rangle = N (p) U(L(\Lambda p)) U(W(\Lambda, p)) |k, \sigma \rangle\;,
\end{equation}
where
\begin{equation}
    W(\Lambda, p) \equiv L^{-1} (\Lambda, p) \Lambda L(p)\;.
\end{equation}
This operator defines the \emph{Wigner rotation}, whose action on $k$ yields
\begin{eqnarray}
     W^{\mu} {}_{\nu} k^{\nu} = k^{\mu}\;,
\end{eqnarray}
i.e., it belongs to a subgroup of the Lorentz group that leaves $k^{\mu}$ invariant. This is the advertised little group. The induced unitary operations of these operators are of the form
\begin{eqnarray}
     U(W) | k, \sigma \rangle = \sum_{\sigma'} D_{\sigma \sigma'} (W) |k, \sigma' \rangle\;,
\end{eqnarray}
where the coefficients $D(W)$ are a representation of the little group. Finally, for any $\Lambda$ we have
\begin{equation}\label{utl}
     U(\Lambda) | p, \sigma \rangle = \frac{N(p)}{N(\Lambda p)} \sum_{\sigma'} D_{\sigma \sigma'} (W(\Lambda,p)) | \Lambda p, \sigma' \rangle\;,
\end{equation}
where $N(p) = \sqrt{k^0/p^0}$, as shown in \cite{Weinberg:1995mt}.

Naturally, the little group depends on the ``choice'' of $k^{\mu}$, particularly on $p^2$ and ${\rm sign}(p^0)$, which are invariant under proper orthochronous Lorentz transformations. We identify two physically meaningful solutions: massive particles, with $p^2 = m^2 >0$ and $k^{\mu} = (m, \bf{0})$; and massless particles, with $p^2 = 0$ and $k^{\mu} = (E, 0,0,E)$. As such, the little group for massive particles is the rotation group in three dimensions, $SO(3)$; and for massless particles is the Euclidean group $ISO(2)$, consisting of rotations and translations in two dimensions. The distinction between the two cases, which will be of importance later on, is that one cannot define a rest frame for photons. For the same reason, the helicity is a Lorentz invariant for massless states and thus only the diagonal term is non trivial in eq. (\ref{utl}), yielding
\begin{equation}\label{uLp}
    U (\Lambda) |p, \lambda \rangle = e^{-i \lambda \alpha(p,\Lambda p)} |\Lambda p, \lambda \rangle\;.
\end{equation}
In other words, a unitary transformation induces just a change of phase, the \emph{Wigner rotation angle}.

Finally, it is crucial to comment on the extension of this concept to a general relativistic framework. Broadly speaking, Einstein's equivalence principle is used to argue that Wigner's formalism applies locally. Then, we can define an orthonormal basis $\hat{e}_{(a)}$ known as a tetrad, representing local inertial frames and for which $g(\hat{e}_{(a)},\hat{e}_{(b)}) = \eta_{ab}$, where $\eta_{ab}$ is the Minkowski metric. This is a convenient basis for the observer, who can now describe through Wigner rotations how moving particle states transform into each other in a curved spacetime \cite{alsing2009, noh2021quantum}.

\subsection{Decoherence}

Now, let us explore how the structure under unitary transformations can lead to different behavior for massive and massless particle states in terms of quantum coherence. We will outline a mechanism for spin decoherence of massive particles by spacetime curvature as studied in \cite{Terashima:2003ds, nasr2008fidelity, Dai:2018vdz}. As a point of detail, notice these studies were also performed in the context of quantum information/communication. 

Consider a wave packet of a spin-1/2 particle, with mass $m$, traveling in a curved spacetime. We define a locally inertial frame as stated before, so magnitudes like the spin can be defined. Then, in such a frame, the one-particle state is described by $|p, \sigma \rangle$, where $p$ denotes the four-momentum of the particle and $\sigma$ its spin along the $z$ axis. Hence, the wave packet can be written as
\begin{eqnarray}
     | \Psi \rangle = \sum_{\sigma} \int d^3 {\bf p}\ N(p^{(a)}) C(p^{a}, \sigma) |p^{(a)}, \sigma \rangle\;.
\end{eqnarray}
The associated density matrix is readily given by $\rho = | \Psi \rangle \langle \Psi |$, but more interestingly, one can define a \emph{reduced density matrix} by integrating over the momenta, yielding
\begin{eqnarray}
     \rho_r (\sigma'; \sigma) & = & \int d^3 {\bf p}\ N(p) \langle p, \sigma' | \rho | p, \sigma \rangle \nonumber \\
     & = & \int d^3 {\bf p}\ N(p) C(p, \sigma') C^*(p, \sigma)\;.
\end{eqnarray}
This expression can describe the state at an initial (proper) time. However, under the influence of a gravitational field the state at a spacetime point $x_f$ will evolve to
\begin{eqnarray}
     | \Psi^f \rangle & = & U(\Lambda(x_f;x_i)) |\Psi^i\rangle \\
     & = & \sum_{\sigma, \sigma'} \int d^3 {\bf p}\ \sqrt{\frac{(\lambda p)^0}{p^0}} C(p, \sigma) D_{\sigma, \sigma'} (W(\Lambda, p)) |\lambda p, \sigma' \rangle \nonumber\;,
\end{eqnarray}
such that the reduced density matrix now reads
\begin{align}
    & \rho_r^f (\sigma'; \sigma) = \sum_{\sigma'', \sigma'''} \int d^3 {\bf p}\ N(p) C(p, \sigma'') C^*(p, \sigma''') \nonumber \\
     & \times D_{\sigma' \sigma''} (W(\Lambda(x_f, x_i), p)) D_{\sigma \sigma'''}^* (W(\Lambda(x_f,x_i),p))\;.
\end{align}
Notice how spin and momentum become entangled through the Wigner rotation. Furthermore, the effects of gravity are also accounted for by the Wigner rotation terms, which contain the information of how the wave packet moves in the curved spacetime (see \cite{Terashima:2003ds, nasr2008fidelity, Dai:2018vdz} for specific technical details). The degree of the entanglement generated by this dynamic can be quantified by the von Neumann entropy,
\begin{eqnarray}
     S = - {\rm Tr} [ \rho_r (\sigma'; \sigma) \log_2 \rho_r (\sigma'; \sigma)]\,,
\end{eqnarray}
which is trivially null for an initial pure state. The crucial point is that entanglement builds up with time. To see this, notice that the reduced density matrix at later times can be written as  
\begin{equation}
\rho_{\mathrm{r}}^{f}\left(\sigma^{\prime} ; \sigma\right)=\frac{1}{2}\left(\begin{array}{cc}
1+\overline{\cos \Omega \tau_{p}} & \overline{\sin \Omega \tau_{p}} \\
\overline{\sin \Omega \tau_{p}} & 1-\overline{\cos \Omega \tau_{p}}
\end{array}\right)\;,
\end{equation}
where the overline denote the average over the momentum distribution (weighted by $N(p) |C(p,\sigma)|^2$). Then, the entropy is given by
\begin{equation}
    S_f = -P \log_2 P - (1-P)\log_2 (1-P)\;,
\end{equation}
where $P = 1/2\ \big[1 - \overline{|\exp(i \Omega \tau_p )|} \big]$. This clearly bounds the entropy as $0 \leq S_f \leq 1$. The authors of \cite{Terashima:2003ds} performed these calculations for a Schwarzschild spacetime, assuming a Gaussian wave packet with spin up along the $z$-direction. They reported a rapid generation of entropy for high velocities for distances relatively close to the Schwarzschild radius $r_S$. This is consistent with a rather small characteristic decoherence time. Naturally, the decoherence time becomes very large when the wave packet is far away from the gravitational source, as well as in the special point $r = 3r_S/2$, where the effects of gravity and acceleration cancel out. Moving on, it was shown in \cite{esfahani2007}, using similar techniques, that similar conclusions can be drawn for more general cases. 

Let us contrast the above with what would happen for the massless case. This proves to be rather trivial, mainly as a consequence of eq. (\ref{uLp}). Indeed, for massless particles the Wigner rotation is a phase change, and there is no entanglement between momentum and helicity as there is between momentum and spin. In fact, using the same technology as above, one has that
\begin{align}
     \rho^f & = | \Psi^f \rangle \langle \Psi^f | \nonumber \\
     & = U(\Lambda(x_f,x_i)) | \Psi^{i} \rangle \langle \Psi^i | U^{\dagger} (\Lambda (x_f,x_i)) \nonumber \\
     & = e^{-i \lambda \alpha} \rho^i e^{i \lambda \alpha} = \rho^i\,,
\end{align}
such that no entanglement would be generated, showing a lack of decoherence as that seen for massive states. 

Nevertheless, a few points are in order. We have discussed only one approach leading to decoherence, where there may be others. In fact, one can prepare in the lab a linear combination of helicities, yielding non-diagonal terms in the density matrix even for the initial states. They can have the form $~\exp(\pm i \phi)$, and after the Wigner rotation, $~\exp(\pm i (\phi + \alpha(\Lambda, {\bf n})))$, where $\alpha$ is the Wigner rotation angle and ${\bf n}$ is the direction of propagation \cite{alsing2009}. However, the latter contribution could still be dismissed upon taking an average over the momentum distribution. 
Notice that there are several instances where observers would measure a zero Wigner rotation angle, strengthening the case for a lack of decoherence. This is the case for observers in a Fermi-Walker frame, i.e., with non-rotating tetrad. Null rotations are also measured for radially infalling photons in a Schwarzschild spacetime, or even for photons traveling in closed loops in the same metric \cite{brodutch2011}, at least for carefully positioned mirrors \cite{Kohlrus:2018cnb}. 

To finish this topic, it is worth mentioning that \cite{Exirifard:2020yuu} estimated the maximum amount of time that a wave packet of massless particles, such as photons, could be under the influence of a gravitational field before it becomes infeasible for the receiver to determine the corrections to the phase that would render the original message. In its most simple form, the condition is $ c t_{\rm max} \ll \ell^2/r_S$, where $\ell$ is the closest distance between the photon and the massive body. This is valid only under certain situations, like that transmitter and receiver are at a similar and very large distance to the massive body and that light deflection is very small. Notice that this above condition is not related with decoherence per se, since it deals with fidelity -- a concept to be discussed below -- but the impossibility of reconstructing the original quantum state resembles decoherence for all intends and purposes. 
Let us exemplify how (un)prohibitive this limit can be. Consider a photon under the influence of the gravitational field generated by the Sun. Taking a distance similar to that between Mercury and the Sun, for which we have $\ell \simeq 6\times 10^7\ {\rm km}$, $r_S \simeq 3\ {\rm km}$, yields the condition $ct_{\rm max} \ll 127\ {\rm ly} \simeq 39\ {\rm pc}$. Thus, the photon could travel a considerable portion of the Milky Way before the limit is violated. Moreover, such distance is comfortably larger than the distance to the closest system of exoplanets, Proxima Centauri, which is at $1.3$ pc from us. This simple exercise reinforces our conclusion regarding gravity not being a decoherence factor for photons. That is not to say that it cannot affect a quantum communication channel, as we have already mentioned and will develop further below. 

\subsection{Fidelity}

Satellite communications have inspired the study of how gravity can affect the quality of a quantum channel. A useful tool to study this problem is the notion of fidelity, which quantifies how close the information received is to the information intended to be delivered. Its functional form is
\begin{equation}
    {\cal F} \equiv {\rm Tr}^2 \left[ \sqrt{\sqrt{\rho} \rho' \sqrt{\rho}} \right]\;,
\end{equation}
where $\rho$ and $\rho'$ are the two density matrices characterizing the systems (or states) under comparison. As an illustrative example it is helpful to consider pure states, say $\rho = | \psi \rangle \langle \psi |$ and $\rho' = | \psi' \rangle \langle \psi' |$, for which the fidelity renders the overlap between the two of them, i.e., $ {\cal F}(\rho, \rho') = \left| \langle \psi | \psi' \rangle \right|^2$. Notice that the fidelity ranges from 0 to 1, and thus expresses the probability that the received state is identified as the one intended to be delivered. 

Clearly, if one searches quantum signals emitted by an ET, the fidelity ideally needs to withhold a value reasonably close to one. In this sense, this is a stronger condition than avoiding decoherence, since the photon can take a state completely different than the initial one (e.g., through a Wigner rotation), while still keeping a quantum coherent state. For example, we could receive a signal and determine that is part of a quantum communication channel, but due to a loss of fidelity, it might be impossible for us to actually determine what was the original message. 

For reference, let us quickly cover some interesting developments regarding gravitational effects on quantum communication. This line of work has been pursued by Bruschi and collaborators (see e.g. \cite{Bruschi:2013sua, Bruschi:2014cma, Kohlrus:2015szb, Bruschi:2021all}). An excellent introduction is given in \cite{Bruschi:2013sua}, where they look at the exchange of single photons in some entanglement distribution protocol considering two observers in a Schwarzschild spacetime. Thus, suppose Bob is on some orbit and sends a photon to Alice, who is on the surface of the Earth. Considering gravitational redshift and other relativistic effects due to the relative motion between the observers, the frequencies at the two places is related by
\begin{equation}
    \Omega_B = \Upsilon\ \Omega_A\;, \qquad  \Upsilon = \sqrt{\frac{1-r_S/r_A}{1-r_S/r_B}}\;,
\end{equation}
where $r_S$ is the Schwarzschild radius of the Earth or any astronomical body in general. An extra factor of $3/2$ on the second term of the denominator can be included to account for additional relativistic effects owing to the relative motion between observers. 

Then, if Alice and Bob observe frequency distributions $f_{\Omega_{i,0}}$ with peak frequencies $\Omega_{i,0}$, these are related by
\begin{eqnarray}
     f_{\Omega_{B,0}}(\Omega_B) = \Upsilon^{1/2} f_{\Omega_{A,0}} \left( \Upsilon \Omega_B \right) \;.
\end{eqnarray}
As mentioned in \cite{Bruschi:2013sua}, given that the transformation above is non-linear, it may be challenging to compensate the effects induced by curvature in realistic implementations. In any case, we can now define a similar measure of the quality of the channel through the overlap $\Delta$ between the two distributions as follows,
\begin{equation}
    \Delta \equiv \int_{0}^{\infty} d \Omega_A f_{\Omega_{A,0}} (\Omega_A) f_{\Omega_{B,0}} (\Omega_A)\;.
\end{equation}
Once again, $\Delta = 1$ characterizes a perfect communication channel, whereas the opposite holds for a null value. Clearly, the fidelity is related to $|\Delta|^2$. Now, taking Gaussian frequency distributions, 
\begin{equation}
    f_{\Omega_0} (\Omega) = \frac{1}{\sqrt[4]{2\pi\sigma^2}} \exp \left( - \frac{(\Omega - \Omega_0)^2}{4\sigma^2} \right)\,,
\end{equation}
and assuming $\Omega_0 \gg \sigma$, so the integration limit can be extended to negative frequencies, one gets 
\begin{eqnarray}
     \Delta = \sqrt{\frac{2(1 - \delta )}{1+(1 - \delta)^2}} \exp \left(- \frac{\delta^2 \Omega_{B,0}^2}{4(1 + (1 - \delta)^2) \sigma^2} \right) \;,
\end{eqnarray}
where $\delta = |\Upsilon^{1/2} - 1|$. Notice that for $r_A\,, r_B \rightarrow \infty$, i.e., in the Minkowski limit, $\delta = 0$ and $\Delta^2 = 1$, as expected in the absence of gravity. Particularly narrow distributions can be obtained through the M\"osbauer isotope $^{57}{\rm Fe}$ and its nuclear transition at $\Omega_0 = 14.4\ {\rm keV}$ and $\sigma \simeq 5\ {\rm neV}$ \cite{vagizov2014coherent, heeg2017spectral, haber2017rabi}, corresponding to X-rays. Then, considering these values and two altitudes given by the surface of the Earth ($r_A = 6371\ {\rm km}$) and the same low Earth orbit examined before ($r_B = 7500\ {\rm km}$), we get $\delta \simeq 5 \times 10^{-11}$ and $\Delta^2 = 0$, which indicates that such a narrow initial distribution would not overlap with the one that arrived. This illustrates our point of the difference between fidelity and decoherence. Indeed, within this context, any sign of the original state was apparently lost, but clearly there is still a quantum signal in the channel. On the contrary, taking other energies available for single-photon sources, like $\Omega_{B,0} = 600\ {\rm THz}$ with $\sigma = 7\ {\rm MHz}$ \cite{wolfgramm2011atom}, we get $\Delta^2 \simeq 0.9999948$, whose value indicates that the quantum communication channel is very good. Finally, it is clear that for stronger gravitational fields and larger trajectories the fidelity will decrease. However, depending on the strength of the gravitational field, the loss of fidelity can be rather small. Case in point, for the same conditions as before, but considering the Sun as the source of the gravitational field, and with $r_A = 1$ AU, one can get as far as $0.01\ {\rm AU} \simeq 1.45 \times 10^6$ km from the transmitter to get the same fidelity, i.e., $\Delta^2 \simeq 0.9999948$. This contrasts the 1129 km ($r_B - r_A$) from the case where the Earth is the gravitational field source. This conclusion is intuitively reasonable, since the gravitational field generated by the Earth is at its surface $g = 9.8\ {\rm m/s^2}$, whereas at 1 AU, the gravitational field exerted by the Sun is $5.9\times 10^{-3}\ {\rm m/s^2}$. The weaker field from the second case explains why a photon would get to travel larger distances while losing the same fidelity than in the first case.
Next, take an unrealistic example where an ET at the same distance as Proxima Centauri b ($r_B = 4 \times 10^{13}\ {\rm km}$) sends a signal to the Solar System, which is received by an astronaut close to Venus ($r_A = 10^8\ {\rm km}$), but where the only gravitational source to be considered in the Sun. Taking the same optical frequency and bandwidth as before, we get $\delta \simeq 7.5 \times 10^{-9}$ and $\Delta^2 \simeq 0.901842$. This indicates a loss of fidelity of roughly $10\%$.

As these estimates indicate, this type of
fidelity loss due to a gravitational field can range from being very severe for X-ray photons to
less significant in the optical band.  Getting back
to our earlier comments, as there is no decoherence here, thus at the receiving end, irrespective of
the fidelity loss, even so severe as for X-rays,
minimally it can still be ascertained that
the received signal contains some kind of
structured quantum communication.  Moreover for
photons of the same frequency they would all
experience exactly the same fidelity loss.  If
one were able to deduce the origin of the signal
and the gravitational field experienced
by the photons throughout their paths, one could
compute the change in fidelity so in principle
adjust the phase shift of the received signal
back to its original sent form.  This supports what
we said earlier, that there would be some
amount of guesswork and trial and error in
understanding the quantum communication signal
that has been received, but all the same one would
at least be aware the signal appears to be from
an intelligent source.

As previously mentioned, there is other work in the literature in this area. One of the most recent ones is \cite{noh2021quantum}, where the Wigner rotation of a photon traveling from the surface of the Earth to a satellite orbit was computed. The authors found that for stationary and free-falling observers with zero angular momentum the Wigner rotation is null. Non-trivial corrections are found whenever the observer has angular momentum. For circular orbits the contribution to the Wigner rotation due to geodetic precession is practically null, but gauge-fixing effects induce a rotation of $~4\times10^{-7}$ for near Earth orbits ($r \sim 300\ {\rm km}$), and $10^{-5}$ for very large distances ($r \sim 10^{11}\ {\rm km}$). For spiraling trajectories the geodetic precession contribution dominates, and the Wigner angle is $\sim 0.0198$ for near Earth orbits and $\sim 0.8229$ for very large distances. For the same trajectory, it is found that gravitation gives a contribution to the quantum bit error rate (QBER) of $3.92 \times 10^{-4}$ for near Earth orbits and $0.537$ for very large distances.

\section{Sources of decoherence}

Having explored the potential effects of spacetime curvature on the feasibility of establishing quantum communication channels to interstellar distances, we shall turn our attention to other factors that could disrupt the quantum state of a photon. A first attempt at this question was reported in \cite{Berera:2020rpl}, where the mean free paths of photons at galactic scales were computed, pointing out that if these are large enough the quantum coherent state of the photon will be sustained. A complementary analysis considering cosmological distances was presented in \cite{Berera:2021xqa}, where the conclusions from \cite{Berera:2020rpl} were reinforced, i.e., X-ray photons are arguably the best alternative for establishing quantum communication at such scales, with the additional point that the transparency window extends up to $z \sim 100$ \cite{Mack:2008nv}. 

In this section we shall look for decoherence factors at smaller scales, which are in some sense, environmental. For this we will focus on the particle population in the Solar System and near Earth as a demonstrative example. Naturally the same conditions will not be exactly the same in other galaxies (or even other regions inside the Milky Way), but this should give us an idea about the factors that an ET civilization may consider if they try to establish contact. Before this, let us quickly review the sources of decoherence pointed out in \cite{Berera:2020rpl}, as those are certainly more generic and are expected to also be representative of other parts of the Universe. 

\subsection{Cross sections, mean free paths and interaction rates}
In the next subsections we will use mean free paths and interaction rates to argue that photons will not (or rarely will) interact with any particle environment. These calculations are possible due to the knowledge of the cross sections of the different possible interactions. Moreover, cross section are the natural quantity to measure experimentally, and thus they are crucial to test any particle physics model. However, these concepts can also be understood from a classical standpoint, and so we will give a quick overview of this with the purpose of covering the basic material required to follow the upcoming subsections. First, consider an incident beam of particles about to hit a target consisting of $N = n A d$ particles, where $n$ denotes the number density of particles, $A$ is the area perpendicular to the incident beam, and $d$ is the thickness. Next, each target particle is assigned an effective area (or cross section) $\sigma$ characterizing the possibility of being hit. Then, the total target area is given by the product $n A d \sigma$, and the probability of an incoming particle to hit one of the targets is 
$$ P = n A d \sigma/A = n d \sigma\;.$$
On average, a particle moves a distance $l$ before interacting, so that
\begin{equation}
    P = 1 = n l \sigma \implies l = \frac{1}{n \sigma}\;.
\end{equation}
Then, for our case, we can define the mean free path $l$ as the average distance a photon travels without interacting. Considering just scattering processes for photons, the interaction rate is readily given by 
\begin{equation}
    \Gamma = \frac{l}{c} = \frac{1}{c n \sigma}\;.
\end{equation}
Finally, from a quantum mechanical perspective, the cross section is given by 
\begin{equation}
    d\sigma = \frac{dP}{t \Phi}\;,
\end{equation}
where $t$ is the time of the experiment, $\Phi$ the flux of particles and $P$ is the quantum mechanical probability of scattering. In turn, this differential probability can be written as
\begin{equation}
    d P = \frac{\left| \langle f | S | i \rangle   \right|^2}{\langle f | f \rangle \langle i | i \rangle} d\Pi\;,
\end{equation}
where $S$ is the so-called $S-$matrix which contains the information about the interactions and thus of the evolution of states in time \cite{schwartz2013}. The initial state is denoted by $| i \rangle$ and the final one by $|f \rangle$, whereas $d\Pi$ denotes a phase space volume of sorts. 

\subsection{Interstellar medium}

Interstellar space has a background distribution of hydrogen, electrons, protons and photons from the cosmic microwave background (CMB), as well as some other heavier elements. For interactions with electrons or protons below the electron mass we need to consider the Thomson cross section, 
\begin{eqnarray}\label{thcs}
     \sigma_{\rm Th} = \frac{8\pi}{3} \frac{\alpha^2}{m^2}\;,
\end{eqnarray}
where $\alpha$ denotes the fine structure constant, and $m$ is the mass of the charged particle. Clearly the dependence on the mass implies that interactions with electrons prevail over interactions with protons. Then, for electron-photon interactions the cross section is $\sigma_{\rm Th} \simeq 6.65 \times 10^{-25}\ {\rm cm}^2$. Using an average number density of free electrons in interstellar space $n_e \approx 1\ {\rm cm}^{-3}$, the mean free path is
\begin{equation}
    l_{\rm Th} = \frac{1}{\sigma_{\rm Th} n_e} \simeq 10^{22}\ {\rm m} \simeq 10^6\ {\rm parsec}\;,
\end{equation}
a distance longer than the size of the Milky Way. A similar analysis can be performed considering the number density of free electrons after reionization, which renders just a $5\%$ probability of interaction. Notice that even that is an over-exaggeration, considering the timeline of galaxy formation and reionization.
On the dense parts on the HII gas region we can take $n_e \simeq 10^4\ {\rm cm}^{-3}$, yielding $l_{\rm Th} \simeq 10^2\ {\rm parsec}$, which is still a considerable distance inside the galaxy, and almost 6 orders of magnitude larger than the distance between Pluto and the Sun. 

Next, we consider the potential interaction of the photon with CMB radiation. Considering energies smaller than the electron mass $m$, the cross section is given by
\begin{equation} \label{EKcs}
    \sigma (\gamma \gamma \rightarrow \gamma \gamma) \equiv \sigma_{\gamma\gamma} = \frac{937 \alpha^4 \omega^6}{10125 \pi m^8}\;,
\end{equation}
where $\omega$ is the energy of the photons in the center-of-momentum frame. The temperature of the CMB is $T_{\rm CMB} = 2 \times 10^{-4}\ {\rm eV}$, so the number density of photons is simply 
\begin{eqnarray}
     n_{\gamma} = \frac{2 \zeta(3)}{\pi^2} T_{\rm CMB}^3 \simeq 411\ {\rm cm}^{-3}\;,
\end{eqnarray}
with average energy density 
\begin{eqnarray}
     \rho_{\gamma} = \frac{\pi^2}{15} T_{\rm CMB}^4 = n_{\gamma} E_{\gamma}\;,
\end{eqnarray}
where $E_{\gamma}$ is the average energy of the CMB photons. Plugging the numerical values given above, we get $E_{\gamma} = 6.34 \times 10^{-4}\ {\rm eV}$. If we consider an X-ray photon with energy 100 keV, we get $\omega \simeq 5.07\ {\rm eV}$ and $\sigma_{\gamma\gamma} \simeq 1.19 \times 10^{-61}\ {\rm cm}^2$, giving a mean free path
\begin{equation}
    l_{\rm CMB} = \frac{1}{\sigma_{\gamma \gamma} n_{\rm CMB}} \simeq 2 \times 10^{58}\ {\rm cm}\;,
\end{equation}
which is much longer than the observable Universe. A more thorough calculation arriving to the same conclusion (lack of interaction of X-rays with CMB radiation) is also shown in \cite{Berera:2021xqa}.

The interstellar medium is dominated by electrons, protons and CMB photons. These quick estimates show that the interactions with this background is negligible (or nonexistent). However, gas and dust are also present in the interstellar medium of the galaxy, with traces of heavier elements. Photons will interact with them through photoabsorption and photoionization. The interstellar medium is transparent for frequencies in the radio and microwave regions \cite{gould1969emission}. At infra-red and visible wavelengths the radiation interacts with atoms in the medium, and thus photons at such energies may not be the ideal (quantum) messengers. Photons in the lower X-ray region ($E\sim 100$ eV) have reasonably large mean free paths, of order 10 parsec, whereas for $10$ keV the mean free path is of order $10^5$ parsec, which is of order of the size of the galaxy
\cite{biswas2013cosmic,cruddace1974opacity}.
Other potentially disruptive factors are the magnetic fields in the galaxy, which can produce a Faraday rotation and also scintillation if magnetohydrodynamic turbulence is present \cite{draine2010physics}. Long wavelength signals are more susceptible to these effects, so X-rays are the prime candidates for quantum communication purposes at such scales. 

\subsection{Local environment}

So far we have not looked at the local environment (at astrophysical scales) of the observers as a factor of decoherence. Certainly this is a more complicated question, and here we can look at our local environment for information, expecting what we extract from it to be common enough to be somewhat similar for other planetary systems. A point in favor of this assumption is that the Sun seems to be a typical star, since it has the same composition as other stars of the same age and at similar positions with respect to the center of their respective galaxies. 

\subsubsection{Matter environment}

First, let us describe the surroundings of the Solar System at different scales. The first environment to be identified, the heliosphere, is (mostly) determined by the activity of the solar wind. This region is immersed in the Local Interstellar Cloud (LIC), a partially ionized diffuse cloud that also plays a role in delimiting the heliosphere. Here, the density of neutral hydrogen is $0.24\ {\rm cm}^{-3}$, whereas the density of electrons is $0.09\ {\rm cm}^{-3}$ and of ionized hydrogen is $0.07\ {\rm cm}^{-3}$. The density of helium atoms is $0.014\ {\rm cm}^{-3}$, with the ratio of ionized to neutral helium of $\sim 45\%$. At a larger scale, our neighborhood is the Local Bubble, a low-density region of roughly 100 parsecs in dimension and which is emptier than the average interstellar medium. It consists mostly of protons, with an average density of $0.005\ {\rm cm}^{-3}$ \cite{balogh2007heliosphere}.

These densities are relatively small in comparison with those near the orbit of the Earth due to solar wind, where the protons and electrons dominate with densities of $\sim 6.6\ {\rm cm}^{-3}$ and $7.1\ {\rm cm}^{-3}$ respectively, and a subdominant contribution from $\alpha-$particles with a density of $0.25\ {\rm cm}^{-3}$ as well as heavier elements \cite{HundhausenA.J1995TSW}. The Sun continually emits these particles with kinetic energies ranging from 1 eV to 10 keV, and their density does not vary significantly within the inner heliosphere. Hence, we can compute the mean free path of photons interacting through Thomson scattering with protons and electrons belonging to the solar wind. Using eq. (\ref{thcs}) and the densities listed above, we get $l \simeq 10^{28}$ m for photon-proton interactions, and $l \simeq 10^{22}$ m for photon-electron interactions. As before, the mean free paths are larger than the size of the Milky Way, even more so for photon-proton interactions. Moreover, notice that for higher energies, where Compton interactions become important, the mean free paths can only be larger, since the Thomson cross section is always greater or equal than that for Compton interactions. However, regardless of the reassuring results, we have yet to analyze the interactions with particles emerging from the most energetic events. Indeed, we need to consider three sources: solar particle events (SPEs), galactic cosmic rays (GCRs) and trapped radiation belts.

 First, SPEs refer to large fluxes of particles emitted during solar flares and mainly from coronal mass ejections. The frequency of these events is correlated to the activity of the Sun. During these events, in particular for coronal mass ejections, the levels of interplanetary particles increase several orders of magnitude over the GCR environment \cite{barth2004space}. The plasma contains ions of every element, but protons dominate with $\sim 96\%$, with peak flux at 1 AU of order $10^5\ {\rm cm}^{-2}\ {\rm s}^{-1}$ and kinetic energies ranging from MeV to a few GeV \cite{barth2004space,bourdarie2008near}. Then, we can estimate an upper limit for the interaction rate simply as follows
 \begin{equation}
     \Gamma_{\rm SPE} = \Phi_{\rm SPE}\ \sigma_{\rm Th}^{p+} \simeq 10^{-26}\ {\rm s}^{-1}\;,
 \end{equation}
 and a mean free path
 \begin{equation}
     l_{\rm SPE} = \frac{c}{\Phi_{\rm SPE} \sigma_{\rm Th}^{p+}} \simeq 10^{34}\ {\rm m}\;,
 \end{equation}
 which is much larger than the size of the observable Universe. Notice that due to the dependence of the cross section on the mass of the proton, as well as the temporary nature of the events, the interactions are even less likely than with solar wind particles. Another reason for this may be that even if the intensities are very large, due to the energy range the number of photons is actually smaller than for the solar wind case. Finally, notice that the abundance of SPEs scale as $R^{-1}$ to $R^{-2}$ for radial distances from the Sun larger than 1 AU, and as $R^{-2}$ to $R^{-3}$ for less than 1 AU. Thus, the interaction rate becomes even smaller in the more external regions of the heliosphere, and does not become significantly larger the more we approach to the Sun. 
 
The situation proves to be somewhat similar with GCRs, whose origin are supernovae explosions (at least partially), and are generated outside of the Solar System. Once again protons dominate the abundances, with $\sim 87\%$, followed by $\sim 12\%$ of $\alpha-$ particles and the rest are heavier ions. The energies can be up to $10^{11}$ GeV, and the flux ranges from $1$ to 10 ${\rm cm}^{-2}\ {\rm s}^{-1}$ \cite{bourdarie2008near}. In consequence, these particles are of no interest for our purposes, since at that energy range other physical processes dominate and even considering Thomson scattering, the flux would lead to interaction rates even smaller than for SPEs. 

Finally, we turn our attention back to our planet. Solar wind particles are trapped by the Earth's magnetosphere, forming a toroidal structure known as Van Allen belts. Two zones are usually identified: the inner belt, where protons and electrons are trapped, and the outer belt, consisting mostly of electrons. Protons carry energies from 1 keV to 300 MeV, whereas electrons in the inner belt have energies ranging from 1 keV to $\sim$ 5 MeV, and in the outer belt from 10 to 100 MeV. Most importantly, the integrated proton fluxes can be of order $10^8\ {\rm cm}^2\ {\rm s}^{-1}$, and $10^7\ {\rm cm}^2\ {\rm s}^{-1}$ for electrons. Clearly, from previous calculations, the mean free path corresponding to photon-proton interactions remains larger than the size of the observable Universe. On the contrary, for photon-electron scattering we have
\begin{equation}
    \Gamma_{\rm VA} = \Phi_{\rm VA} \sigma_{\rm Th}^{e-} \simeq 10^{-18}\ {\rm s}^{-1}\;, 
\end{equation}
yielding a mean free path
\begin{equation}
    l_{\rm VA} = \frac{c}{\Phi_{\rm VA} \sigma_{\rm Th}^{e-}} \simeq 10^{25}\ {\rm m} \simeq 10^{8}\ {\rm parsec}\;.
\end{equation}
In consequence, the photon could cross through the belts without interactions. 

\subsubsection{Radiation environment}

The final environmental factor to be considered is the photon population in the Solar System. Naturally, this is mostly determined by the Sun itself, which is near black body radiation at an effective temperature $T \simeq 5800$ K, more or less between 0.2 and 3 $\mu$m. There is also some extreme UV radiation coming from active regions, as well as X-rays, where the background dominates over the solar output. The X-ray background is thought to be produced by accretion disks around active galactic nuclei \cite{hill2018spectrum}. 

For the sake of concreteness, we shall take data from the Solar Irradiance Reference Spectra (SIRS) \cite{chamberlin2009new} containing the irradiance spectra from 0.1 nm to 2400 nm in the heliosphere for three periods of low solar activity. From this, we can estimate the flux of photons on a region between two wavelengths as follows 
\begin{eqnarray}
    \Phi = \int_{\lambda_1}^{\lambda_2} \frac{F_{\lambda}}{E_{\lambda}} d\lambda\;,
\end{eqnarray}
where $F_{\lambda}$ is the irradiance and $E_{\lambda} = hc/\lambda$. Then, the interaction rate of our test photon can be readily computed as
\begin{align}\label{rate1}
    \Gamma & = \int_{\lambda_1}^{\lambda_2} \sigma_{\gamma \gamma}\frac{F_{\lambda}}{E_{\lambda}} d\lambda \nonumber \\
    & = \int_{\lambda_1}^{\lambda_2} \frac{938 
    \alpha^4}{10125\pi} \frac{5}{16} \frac{E_{\rm test}^3 E_{\lambda}^3}{m_e^8} \frac{F_{\lambda}}{E_{\lambda}} d\lambda\;,
\end{align}
where we have used the photon-photon interaction cross section formula (eq. (\ref{EKcs})), and $E_{\rm test}$ denotes the energy of the photon establishing the quantum communication channel. Performing the numerical integration over the available wavelengths, and taking $E_{\rm test} = 100$ keV, we obtain $\Gamma \simeq 7 \times 10^{-33}\ {\rm s}^{-1}$. This corresponds to a mean free path $l \simeq 10^{40}$ m, which again, is much larger than the size of the observable Universe. 

Next, let us quickly look at the region of the spectrum not covered above, the X-ray background. As pointed out in \cite{ajello2008cosmic,hill2018spectrum}, a double power law of the form
\begin{equation}
    \frac{dN}{dE} = \frac{A}{(E/E_b)^{a_1} + (E/E_b)^{a_2}} [{\rm photons}/({\rm cm}^{2}\ {\rm s}\ {\rm sr}\ {\rm keV})]
\end{equation}
fits the spectrum rather nicely, particularly in the energy range from 2 keV to 2 MeV. The best-fit parameters reported in \cite{ajello2008cosmic} were $A = 10.15 \times 10^{-2}$, $E_b = 29.99\ {\rm keV}$, $a_1 = 1.32$ and $a_2 = 2.88$. Then, the interaction rate is computed as follows
\begin{equation}
    \Gamma_{\rm XR} = \int \int \sigma_{\gamma \gamma} \frac{dN}{dE} \cos \theta\ dE d\Omega\;.
\end{equation}
Setting $E_{\rm test} = 100$ keV and the electron mass as the upper energy limit to remain within the validity regime of the cross section formula, the numerical integration yields $\Gamma_{\rm RX} \simeq 8 \times 10^{-52}\ {\rm s}^{-1}$, with a mean free path $l_{\rm RX} \simeq 10^{59}$ m. Clearly, the possibility of interaction between a photon in the X-ray range and the CXR is basically null. Furthermore, the situation is even better for test photons in the optical range, where for a typical energy of 10 eV, the mean free paths would increase by a factor of $10^4$.

A similar process can be followed for the optical band of the extragalactic background light (EBL), which consists of photons emitted by stars and other astrophysical objects during the cosmic history. The optical and near-IR bands there are dominated by direct stellar emission, whereas the far-IR is dominated by stellar radiation reprocessed by dust in the galaxies \cite{hauser2001cosmic, singh2020extragalactic}. The spectral energy distribution of the optical band can be (crudely) approximated by a quadratic polynomial of the frequency in a logarithmic scale, i.e., $\ln (\nu I_{\nu}) \simeq c_0 + c_1 \ln \nu + c_2 \ln^2 \nu$. \footnote{For reference, $I_{\nu}$ is the Planck intensity function for black body radiation. The constants are $c_0 = -111.231$, $c_1 = 15.2089$ and $c_2 = -0.623$, which gives a spectral energy distribution in units of ${\rm nW}\ {\rm m}^{-2}\ {\rm sr}^{-1}$ with $\nu$ given in GHz.} Integrating this over the logarithm of the energy gives a brightness of $21\ {\rm nW}\ {\rm m}^{-2}\ {\rm sr}^{-1}$, which is close to the value reported in \cite{dole2006cosmic} of $23\ {\rm nW}\ {\rm m}^{-2}\ {\rm sr}^{-1}$ and well within observational uncertainties. This gives some credence to the interaction rates to be computed from this approximation. The interaction rates are thus given by
\begin{align}
    \Gamma_{\rm Opt}^{\rm EBL} = \int_{\nu_1}^{\nu_2} \sigma_{\gamma \gamma} \frac{I_{\nu}}{E_{\nu}} d\nu\;,
\end{align}
where the integration was performed from $\nu = 4\times 10^4$ GHz to $10^6$ GHz. Once again, setting $E_{\rm test}=100$ keV, we obtain an interaction rate $\Gamma \simeq 5\times 10^{-44}\ {\rm s}^{-1}$, with a mean free path $l \simeq 6\times 10^{51}$ m. Notice that EBL radiation at these frequencies is subdominant inside the Solar System (which explains why the mean free path here is larger than the $10^{40}$ m obtained from eq. \eqref{rate1}), but can nevertheless dominate outside the heliosphere or in the interstellar medium. Similar results can be expected in the IR, where the total brightness is $24\ {\rm nW}\ {\rm m}^{-2}\ {\rm sr}^{-1}$. This, together with the lower cross sections $\sigma_{\gamma \gamma}$ in the IR implies that the possibility of interactions with photons at these energies remains basically null. 

\section{Communication with an unknown distant civilization through quantum teleportation}

\subsection{Signature of a quantum teleportation signal}

Establishing contact with an unknown distant civilization would be challenging through any means of communication since no common language is shared. Thus, the receiving party will
need to decipher any such message if it were received.  There is also the question of how much
information is the sender attempting to communicate.
Then, the most basic goal in establishing such a communication channel is to unambiguously convey to the receiver that the signal is coming from an intelligent source and for it not to be easily confused with any natural background sources. In this respect, quantum teleportation would have two correlated signals, and that alone would be indicative as being sent deliberately by an intelligent source. To make any further sense of such a signal would require a degree of guesswork. The sender would realize this and so would most likely repeatedly send the same message, so that the receiver had many opportunities to test it and make sense of it.

How exactly would a quantum teleportation signal appear at the receiving side? Any quantum channel (mediated by photons) would appear as a series of single photons, if they were qubits, or bunches of photons, if they were higher dimensional entangled states. For quantum teleportation, it would be two electromagnetic signals that arrive at the same time and from the same direction in the sky. The photon part of the shared entanglement for the simplest form of quantum teleportation, given in Section \ref{qtssect}, would just appear as a monochromatic, unpolarized electromagnetic signal.  For more complex forms of teleportation, it is possible there would be photons at different frequencies. Other than that, if observed classically, there would be no obvious correlation or structure to the signal. All the information in the entangled photons is encoded within their quantum state, in terms of their polarization quantum numbers and perhaps other observables such as frequency and angular momentum state. Thus, if one were not observing the signal properly as a quantum state, all the quantum information would simply be lost.  The other signal would be the classical channel, and that would have more structure to it. For the simple example in Section \ref{qtssect}, the classical channel only requires four different types of signals. For more complicated quantum teleportation protocols, this number may change, but nevertheless what would be observed at the receiving end would be some finite number of signals that are ranged over. If such a signal profile was found, this might suggest to the receiver that the joint two-signal transmission was an attempted quantum teleportation.  In that case the receiver would know the other signal would be part of the shared entanglement and so try and measure it using proper techniques to treat its quantum properties.

The Bell states are maximally entangled states between two two-state systems.  One could assume that these would be standard and understood by any advanced civilization. Thus if the receiver believed they were detecting a teleportation signal, the step they may take is to assume the photons in the quantum states are one part of a shared entangled Bell states, with the other being with the sender. In this case the receiver would not know what specific entangled Bell state they were receiving a part of.  However there would be a choice of only four. Thus the receiver could simply guess and go through all possible combinations and see for which the received signals made any sense. This is one reason why it would be important for the sender to send the exact same message repeatedly, so as to allow the receiving end to test each of the many repeated signals through this guesswork process until a outcome was attained that looked the most sensible.

Finally, notice that the example given in Section \ref{qtssect} illustrates the teleportation of a single qubit state. This can be extended to teleportation of multi-particle and multiple degrees of freedom, thus allowing for more information to be communicated. As quantum teleportation is a linear operation on quantum states, in principle this should be extendable to multiple degrees of freedom.  In practice, this poses challenges but progress is being made. The first step is to create higher dimensional entangled states by combining many two-dimensional qubit states or utilizing also other degrees of freedom. Simultaneous entanglement could be done between multiple degrees of freedom of a particle, which could mean entanglement involving polarization, orbital angular momentum, frequency, and wavenumber.  There are developments in creating such states \cite{deng2017quantum}, including in \cite{wang2015quantum} where quantum teleportation was shown for a composite quantum state of a single photon possessing both spin and angular momentum. Other examples of higher dimensional entanglement include utilizing orbital angular momentum to create a qutrit (three dimensional entanglement) \cite{vaziri2002experimental} and twelve dimensional state \cite{dada2011experimental}. Orbital angular momentum has also been combined with a radial quantum number to create a entangled state above a hundred dimensions \cite{krenn2014generation}. Multiparticle entangled states are also of interest, with a three--photon entangled state created in \cite{malik2016multi}. This is a developing area, but teleportation of multiple degrees beyond just one qubit is viable. Alongside this experimental progress, more sophisticated quantum teleportation protocols have developed with multi--qubit states. Teleporation protocols with three qubit Greenberger-Horne-Zeilinger (GHZ) states has been considered in \cite{ghosh2002entanglement, hsu2014quantum}. Higher qubit teleportation protocols also have been developed such as involving five qubits \cite{saha2012n} and N--quibits \cite{liu2007quantum}.

\subsection{Advantages to quantum teleportation protocols of communication}

Beyond just providing an unambiguous indicator of an intelligent source sending the signal, a quantum teleported signal may also allow considerable information transfer, and this could be the main argument in favor of this communication mode. Communication with extraterrestrial intelligence (CETI) also
known as exosemiotics is a field that examines types of communication signals to and from
extraterrestrial intelligence \cite{Reed2000}. Such studies have focused on classical communication
and amongst other approaches considered mathematical and computational forms of communication.  Quantum teleportation offers new modes of communication along these lines, possibly utilizing quantum computing or some type of quantum communication protocol.

Quantum teleportation allows for
information transfer in a different way from classical communication,
which has some beneficial consequences.
An $n-$qubit wavefunction has $2^n$ possible states. Thus a quantum
wavefunction comprised on $n$ qubits
in principle could contain a linear combination of all
these $2^n$ states.  At the face of it, it would seem such a wavefunction could
contain a huge amount of information. However recall that in
quantum mechanics no information is acquired from the quantum
state vector until a measurement is done, at which point
the state vector collapses to one possible state and all
other information in the state vector at that point is lost.
A theorem by Holevo states that by communicating $n$ qubits, one cannot
transmit more than $n$ bits of information \cite{Holevo73}.
Thus a direct use of teleportation for transferring of bits of
information cannot enhance the amount of information
transferred.  However if the wavefunction of the received
quantum channel remains unmeasured,
it is possible it can be manipulated by particular quantum
operators, and through such an operation a more significant
use of the wavefunction for treating information can arise.
This is precisely what researchers today are actively
trying to understand in the field of quantum computing.
One of the standard applications understood
for quantum teleportation is for communication between
quantum computers or different
parts of a single quantum computer.
Quantum teleportation allows for a quantum state to be transfered
without losing its quantum coherence.  Thus if a distant
signal was in fact a quantum teleportation signal, 
rather than immediately measuring the quantum state,
it may
need to be received by a quantum computer. At present quantum computing is in a early developmental stage, but there are some indications of its potential to
process information in certain cases far more efficiently
than with classical
computing \cite{aaronson2008}.  These possibilities
could be the very reasons why quantum teleportation offers the best
approach for communication between civilizations separated
at interstellar distances. 

In the context of information processing,
there are some examples where quantum computing has shown an
advantage to classical computing.  For example
the Grover algorithm \cite{grover96},
allows for the sorting of $N$ items quantum mechanically in
$O(\sqrt{N})$ operations whereas classically
it requires $O(N)$ operations.
Another example is the Shor algorithm \cite{shor99}
where quantum computing can factor an integer
$N$ in polynomial time in $\ln N$, which is exponentially
faster than any classical factoring algorithm.
There is also the concept of quantum supremacy, 
that certain problems quantum
computers can solve which a classical computer in all
practical terms can not.
This is an active area of research with possible
claims to success by researchers at Google \cite{aruteetal2019} and at USTC \cite{zhong20, zhong21, wu21}.

From information theory it is known that the
von Neumann entropy
allows greater compression of quantum information
than the Shannon entropy allows for 
classical information \cite{schumacher95}.
Although the technology to exploit this difference is yet to be developed, what is suggestive
is that greater information  compression may be possible in quantum versus classical communication.
There are some known cases where
communication using a quantum protocol has 
advantages
to a classical protocol. For example, for certain communication
complexity\footnote{Communication complexity is the idea of how many bits of information two
parties need to exchange to solve a specified computational 
task \cite{yao79, KW90}.} problems it has
been shown that quantum communication can be exponentially
faster to classical communication \cite{Raz99}.
One-way communication complexity is a special case
of this idea where the sender communicates a single message to
the receiver, after which the receiver then computes
the desired output.  This would be the class of problems
relevant to interstellar communication by an intelligent
source.  For certain mathematical models it has
been shown that one-way communication complexity
also is exponentially faster for a quantum protocol 
to any classical one \cite{bar2004,regev2011}.
At present these ideas from quantum information theory
alongside quantum communication are
developing areas of study. However what already is
understood is there are  some quantum advantages and
perhaps they can be utilized for
communicating significant information
more efficiently than by any classical method.
That boost in efficiency would be very useful
to two distant civilizations, where information exchange
comes at a premium due to the long transmission time. Indeed, communication with a distant intelligent civilization would involve sending
signals that may take years to reach their destination. Therefore,
an advanced civilization may deem it is only worthwhile
to make contact with a distant civilization if the
communication mode is efficient and
contains as much information as possible.
With this in mind, quantum teleportation certainly offers some advantages.

Understanding the utility and potential of
quantum information and how it can be processed
in quantum computing
is at the moment still a developing subject.
At present there is no definitive process 
one can point to in quantum computing that would
make it an attractive argument for quantum teleportation
as the preferred communication approach
between distant civilizations.  For now it is simply
a distinctively different alternative with its
own specific advantages.

\section{Discussion}

The program for the search for extraterrestrial intelligence (SETI) has focused on looking for ``classical signals'' in the radio and optical ranges, following the guidelines set by the seminal papers of Cocconi and Morrison \cite{morrison1959searching}, and Schwartz and Townes \cite{townes1961interstellar}. In \cite{Berera:2020rpl} a fundamental observation was made that photons in certain frequency bands can maintain quantum coherence over interstellar distances, opening the way for the proposal that quantum communication is possible.
This paper has analyzed the different aspects to be considered in order to establish a quantum communication channel across interstellar distances, as well as the challenges of detecting and interpreting such signals. An integral part of this study has been the identification of the sources that can disrupt the quantum state of photons sent as part of the communication protocol. 

By looking at the concept of Wigner rotation and focusing on a particular form of entanglement generation, we made the case that for photons, there would be no decoherence induced by gravitational fields. It must be emphasized that this is only the case for photons, as qubit implementations through the spin of electrons are expected to lose coherence due to a spin-curvature coupling. We have also strengthened our case by pointing out specific cases where the Wigner rotation angle is null under the influence of gravity, as shown in \cite{brodutch2011,Kohlrus:2018cnb}, as well as using other criteria proposed in \cite{Exirifard:2020yuu}, where the maximum time of gravitational influence on a photon that would allow to recover the message was proposed. We showed how this limit can be rather non-restrictive in comparison to the time scales required to propagate the communication process inside the galaxy. 

We also reviewed some of the existing literature on the influence of gravitation on fidelity, which has been widely studied in the context of Earth-to-satellite communication. This was particularly useful to stress the difference between fidelity and decoherence. Indeed, fidelity quantifies the goodness of the communication channel, so the lack of decoherence is necessary but not sufficient for establishing a viable quantum channel. Notice that this statement is true for the kind of setup we looked at. For terrestrial communications, there are different ways to sidestep noise effects, like through qubit transduction \cite{lim2015experimental} or by establishing a network of quantum repeaters \cite{ruihong2019research}. However, in order to establish communication across astronomical distances, we have to at least expect decoherence not to be an issue since fidelity changes due to gravitational effects are certainly expected. In fact, for Gaussian pulses, as mentioned
already, in the X-ray range one could expect a complete loss of fidelity already at low Earth orbits, whereas the quantum coherence remains intact. For other regions of the spectrum accessible for single-photon sources, like in the optical or near the UV, the loss of fidelity is not nearly as bad, with an estimated loss of 10\% for distances similar to that between Proxima Centauri and the Solar System.  For any of these cases,
as we already discussed, in the event of minimal decoherence, the effects of fidelity loss 
from a gravitational field in principle could
be mitigated.  All things considered, 
it is clear the receiver end of the channel will need to infer many of the initial conditions of the signal in order to reconstruct the original message, but regardless of the plausibility of this process, they could at least in principle determine the quantum nature of the signal, thus the 
possibility of it emerging from an intelligent 
source. 

Considering this, we have explored the potential sources of decoherence across the interstellar medium, where it has been shown before that X-rays (we restricted
our study only to energies below the electron mass) have mean free paths long enough to even cross galactic distances without interactions, and the same applies for radiation at radio frequencies and up to the microwave region \cite{Berera:2020rpl}. In this work we extended this analysis by looking at more local environments taking the heliosphere as an example. The mean free paths associated with interactions between the photon and particles emitted during solar winds, solar particle events, and galactic cosmic rays are at least larger than galactic scales, and sometimes larger than the observable Universe. The same result was found when looking at the particle content of the Van Allen belts of the Earth. Finally, we also checked the photon content in the heliosphere covering the entire solar spectrum in addition to the X-ray content from the cosmic background, leading to the same conclusions as before. 

In light of these results, it is plausible that quantum communication mediated by photons could be established across interstellar distances, in particular for photons in the X-ray region below the electron mass. That is not to say that radio signals or in the optical range are out of the question, as they also show fairly large mean free paths. This is a remarkable result, since in principle one has to assume there is very little chance to observe such a quantum communication signal arriving from interstellar distance.  Nevertheless this paper along with our previous ones \cite{Berera:2020rpl,Berera:2021xqa} have demonstrated that such a signal can propagate through large interstellar distances without decoherence. As such, this is a viable signal for astronomy. In any case, dealing with the loss of fidelity, as low as it may be, is certainly one of the biggest challenges. From the technological side, photonic qubits are usually in the optic regime, as already mentioned. Nevertheless, it has been pointed out that X-rays are particularly appealing for quantum information purposes, as the detection efficiency is higher than for optical photons, are more robust, have more penetration power and better focusing capability, making them more suitable to be information carriers. It seems that the best way to generate such photons is through the ${}^{57}{\rm Fe}$ isotope, which shows a magnetic dipole resonance at $14.4$ keV and a linewidth of 4.7 neV \cite{haber2017rabi, gunst2016logical, vagizov2014coherent, heeg2017spectral}. Thus, almost by happenstance the X-ray region provides us the better capabilities for terrestrial and astronomical quantum communication applications. 

In fact, the future potential of X-rays for classical communication is already recognized for space communication due to its improved beam collimation and greater information transfer, and is being tested with the NASA XCOM satellite experiment \cite{nasa}. In a separate direction, entanglement with X-ray photons is a growing area of research interest \cite{ispirian2009production,shwartz2011polarization,borodin2016x,sofer2019quantum}. However, atmospheric absorption may require that signal receival would need to be space based. On the contrary, the atmosphere is transparent to radiation at optical wavelengths, as well as for wavelengths in the order of a centimeter. Observatories like ALMA can also conduct observations for wavelengths of the order of a millimeter, which may suggest that this is the way to go for Earth-based searches. In fact, radio astronomy observations require an array of antennas in order to attain angular resolutions that allow to resolve a single object. Thus, the signals captured by each antenna in the array need to be combined using interferometry \cite{pannuti2020physical}, and one could speculate that similar principles could be used in order to determine the quantum nature of a signal. Nevertheless, other technical difficulties like the need to amplify the signals using photomultiplier tubes may be insurmountable, since that process would effectively destroy the quantum state of the arriving photons. In that sense, X-rays may be the best alternative for interstellar quantum communication, although the optical band is also worth exploring in spite of the smaller mean free paths. In \cite{hippke2021searching}, following on the observation in \cite{Berera:2020rpl} for intelligent quantum communication signals to be possible over interstellar distances, a series of tests using available technology were proposed to look for intelligent quantum signals. As such, these could mainly exploit optical wavelenghts for which, as already said, there is limited distance in the interstellar medium for quantum coherence of the photons to be maintained.

Next, we commented on some of the ways we could expect to identify a quantum signal coming from space, particularly from an intelligent civilization, and the advantages of establishing such a channel over classical communication. We have used quantum teleportation as a simple yet illustrative example to argue how it could be a likely choice for an extraterrestrial quantum signal. Naturally, there are other protocols of quantum communication, all of which take advantage of the properties of quantum entanglement, so the main ideas explored here can also be applied. Some of these protocols include quantum key distribution \cite{bennett2020quantum}, quantum secure direct communication, superdense coding \cite{bennett1992communication}, as well as variants of quantum teleportation such as remote state preparation.

In conclusion, taking into account the decoherence of photons from gravity and electromagnetic interactions in the interstellar and Solar System medium, we have found that a quantum communication signal can traverse vast distances, where interactions with the interstellar medium are far more likely. Our study has found that the distance a photon can propagate without noticeable decoherence in the X-ray band is up to 1 MPc in the interstellar medium, although it can be reduced to 100 pc for dense HII gas regions. Our conclusions concerning gravitational effects and local environments can also be extended to the optical and radio/microwave bands. Hence, interstellar extinction (scattering and absorption by dust and gas) would be the main decoherent factor of photons in the optical band. 
On the contrary, the interstellar medium is quite transparent to radio/microwave photons, but there could be absorption corresponding to the 21 cm line, although this is more likely against a continuum radiation background. Then, photons in this band could also travel considerable distances before interacting (if at all). 
Notice that the Kepler space mission and other astronomical searches for Earth-like planets have identified tens of terrestrial planets in solar systems within just tens of parsecs distance from Earth \cite{exo}. Thus, many Earth-like planets are close enough for a quantum teleportation signal to reach without significant degradation.

Looking more broadly at quantum communication, one could ask why an extraterrestrial quantum teleported signal might be favored over any classical approach or other quantum communication protocols. The argument may be as simple as that is the available technology and it is less accessible to develop, for example, what might already be outdated radio communication apparatus (alien scientists may have the same funding challenges as their counterparts here on Earth). Another argument in favor of quantum teleportation over other quantum communication modes is that it would come as two related signals and could provide a better signature for detection. When measured in a classical way, a beam of quantum correlated photons sent on its own would lose any quantum aspect of the signal. Thus, receiving a dual signal may be more indicative to the receiver to properly analyze the incoming signal and look for its quantum features. 
If such features were found, it would remove any doubt that the signal emerged from an intelligent source. Moreover, more complex quantum teleportation protocols to the illustrative example in this paper could send multi-photon states, which themselves could be in quantum correlated states, including, for example, Bell states \cite{saha2012n}. An advanced civilization attempting to achieve first contact with another civilization would want to send a signal that was readily identifiable, even universally understood. In the subject of quantum entanglement, the Bell states have achieved that status of being universally identifiable. Thus they offer a viable option for teleportation to an unknown civilization.

\begin{acknowledgments}

We wish to thank Alan Heavens and Yvonne Unruh for helpful information regarding the particle content in the Solar System. AB also thanks Michael Hippke for helpful discussions. AB is partially funded by STFC. JCF is supported by the Secretary of Higher Education, Science, Technology and Innovation of Ecuador (SENESCYT).  For the purpose of open access, the authors have applied a Creative Commons Attribution (CC BY) licence to 
any Author Accepted Manuscript version arising from this submission.
\end{acknowledgments}

\bibliography{apssamp}

\end{document}